\documentclass[12pt]{iopart}
\usepackage{epsfig,bm}
\usepackage{graphicx,cite}
\usepackage{amssymb,psfrag,textcomp}
\usepackage{amsthm,upgreek}

\newcommand{\be}{\begin{equation}}
\newcommand{\ee}{\end{equation}}
\newcommand{\bd}{\begin{displaymath}}
\newcommand{\ed}{\end{displaymath}}

\begin{document}
\title[The man-mosquito problem in the presence of drift]{On the  passage properties of the gradual capture of a diffusive particle in the presence of drift}
\author{Javier  Quetzalc\'oatl Toledo-Mar\'in}
\address{Departamento de F\'isica-Qu\'imica, Instituto de F\'isica, UNAM, P.O. Box 20-364, 01000 M\'exico, D.F., M\'exico}
\author{Isaac P\'erez Castillo}
\address{Departamento de Sistemas Complejos, Instituto de F\'isica, UNAM, P.O. Box 20-364, 01000 M\'exico, D.F., M\'exico}
\ead{isaacpc@fisica.unam.mx}
\begin{abstract}
We investigate a  stochastic process consisting  of a  two-dimensional  particle with anisotropic diffusion, mortality rate and  a drift velocity, in the presence of an absorbing boundary. After the particle has encountered the boundary, the process is restarted with updated values of its diffusion constants and drift velocity. We then derive the joint probability that, after $M$ encounters, the particle is absorbed at a point  of the boundary  at a given time and show that, under certain conditions, the  eventual hitting probability corresponds to a Bessel distribution. In the context of the man-mosquitoes problem, the mosquito is able to gradually capture the man, after which the mosquito follows a diffusion process with no drift. Our results are compared thoroughly with Monte Carlo simulations showing excellent agreement.
\end{abstract}
\maketitle
\section{Introduction}
Random walks and Brownian motion are simple approaches used to model successfully  a wide variety of processes ranging from biology, to chemistry, to physics, to economy, and sociology. In particular, first passage processes, stochastic processes that stop after certain event happens, have attracted considerable research in different fields such as, to name a few, phase transitions, coagulation phenomena, neuron dynamics,  and chemical reactions \cite{ben2003ordering, weiss1994aspects, meerson2015mortality, bray2013persistence, bramsoncapture, kesten1992absorption, li2002normal, katori2002scaling, fisher1984walks, redner1984kinetics, blumen1984target, bendler1988first, redner2001guide, bramson1991asymptotic, fisher1988reunions, krapivsky1996kinetics, redner1999capture, redner2014gradual,salminen1988first, turban1992anisotropic, igloi1993inhomogeneous, krapivsky1996life, grabiner1999brownian, benichou2012non, klass1983maximum, gabel2012can}.\\
Some of these first passage processes correspond to models in which an ensemble  of one-dimensional particles follows a diffusive process until two or more particles meet and subsequently the process stops.  For instance, Fisher and Gelfand \cite{fisher1988reunions} studied the behaviour of three vicious random walkers which diffuse freely in one dimension with arbitrary diffusivities, obtaining the full distribution in the continuum limit.  In  \cite{bramsoncapture} the authors studied the capture time for $N$ predators and a prey and showed 
that the average capture time diverges for up to three predators, but for four or more predators it becomes finite. In subsequents works it was proven that for large $N$ the mean capture time remains finite \cite{kesten1992absorption,li2002normal}.\\
In a fairly elegant setup of the problem, Krapivsky and Redner \cite{krapivsky1996kinetics, redner1999capture} 
studied the survival probability of the so-called lamb-lions problem, in which a one-dimensional diffusing prey -the lamb- is in proximity to $N$  one-dimensional diffusing predators -the lions- and showed that when the lions are all to one side of the lamb, the survival probability decays as a non-universal power law with an anomalouos exponent proportional to $\log(N)$ . They also showed that, in this initial setup, the best survival strategy for the lamb is to diffuse faster than the lions. In contrast, for a two-sided system, in which the lions initially surround the lamb, the best survival strategy for the lamb is to remain still. \\
Further properties of the lamb-lions problem were further studied in \cite{ben2003ordering} in what is called the leader and laggard problems. Here the authors had $N$ diffusive particles and studied the probability that the leftmost particle remains the leftmost particle (the leader problem) and the probability that the rightmost particle never becomes the leftmost one (the laggard problem). For 3 particles they found an anomalous exponent for the decay of the survival probability in agreement with that of  \cite{redner1999capture}, and a precise estimate for four particles. Although the problem for a large finite number of particles remains elusive, when taking the limit of an infinite number of particles, one can map the problem to that of a diffusive particle in the presence of an approaching absorbing  boundary whose position goes as $\sqrt{t}$ \cite{turban1992anisotropic, igloi1993inhomogeneous, krapivsky1996life}.  This approach to the problem yielded an anomalous exponent for the decay of the survival probability  of $\frac{1}{4}\log(N)$ for large but finite number of particles, and the survival probability as $\sim e^{-\log^2(t)}$ for an infinite number of particles \cite{redner1999capture, krapivsky1996kinetics}.\\
Inspired by the lamb-lions problem  and motivated by the unusual phenomenology of locally activated random walks \cite{benichou2012non}, Redner and B\'enichou  fairly recently introduced the so-called  man-mosquitoes problem. Here,  independent mosquitoes, each with fixed diffusivity, repeatedly bite a diffusing man. After each bite of the mosquito, the diffusivity of the man is reduced by a fixed amount, while the diffusivity of each mosquito is unchanged, and  the mosquito immediately moves equiprobably either a distance $+a$ or $ -a$ from the man after each encounter, starting the process afresh. After the man has been bitten $M$ times,  the authors assumed that his diffusion constant is zero, that is the man is dead, and studied the position and time elapsed  at which the man dies \cite{redner2014gradual}. They found, for instance, that since the eventual hitting probability is known to be a Cauchy distribution, which is stable under convolution, the position at which the man dies after $M$ bites also follows a Cauchy distribution  \cite{redner2014gradual}.\\
The main goal of the present work is to generalise the study done in \cite{redner2014gradual} when to  the man-mosquitoes problem we add  a velocity drift and a mortality rate.  In particular, we focus on understanding under which conditions the statistics of the  process  after $M$ encounters is given by stable distribution.  This paper is organised as follows: in \sref{sect:md}, we introduce the model and solve it rather straightforwardly using the method of images in the original coordinate system. We show that the eventual hitting probability, or the probability to find the man at position $x$ after the first encounter, follows a Bessel distribution or also called a normal inverse Gaussian distribution. In \sref{sect:ap} we investigate under which conditions the probability of finding the man at position $x$ after $M$ encounters also follows a Bessel distribution. Under these  conditions we see that the mosquito is able to capture the man, while at the same time the mosquito becomes satiated. Our results are thoroughly checked by means of Monte Carlo simulations in \sref{sect:mcs}. Finally, \sref{sect:c} is for conclusions.
\section{Model definitions}
\label{sect:md}
In the context of the man-mosquitoes problem the process is introduced as follows: suppose we have a one-dimensional desperate and diffusing man with diffusion constant $D_1$ and drift $v_1$. A one-dimensional hungry and diffusing mosquito is chasing him with diffusion constant $D_2$ and  drift $v_2$. When they encounter the mosquito bites the man, changing his health and possibly also changing the state of the mosquito. The process is then restarted afresh with a different set of parameters and repeated $M$ times, that is, the mosquito bites the man $M$ times. It may also happen that the process dies before anything happen, so we include a mortality rate.\\ 
This process can also be more generally presented as follows: consider a two-dimensional particle with position coordinate $\bm{x}=(x_1,x_2)\in\mathbb{R}^2$ with anisotropic diffusion $(D_1,D_2)$ and in the presence of a velocity drift $\bm{v}=(v_1,v_2)$ and mortality rate $\varrho$. Its diffusion equation is
\begin{equation}
\frac{\partial P(\bm{x},t)}{\partial t}+\varrho P(\bm{x},t)+\sum_{i=1}^2v_i\frac{\partial}{\partial x_i}P(\bm{x},t)=\sum_{i=1}^2D_i\frac{\partial^2}{\partial x^2_i}P(\bm{x},t)\,.
\end{equation}
We seek the solution of this problem with initial conditions $x_i(t=0)=x_{i0}$ for $i=1,2$  such that, without loss of generality,  $x_{20}>x_{10}$ and with an absorbing boundary/line $\mathcal{H}_{x}=\{\bm{x}\in\mathbb{R}^2 | x_1=x_2=x\}$. The standard way to tackle the problem is firstly to do a change of variables in which the system is isotropic and then do a rotation so that the absorbing boundary corresponds to one of the new axis (see, for instance, \cite{redner2014gradual}). We notice instead that one can fairly straightfowardly use the method of the images to write a simple expression for the occupation probability $P(\bm{x},t)$ with an absorbing boundary at $\mathcal{H}_x$, \textit{viz}.
\begin{equation}
P(\bm{x},t)=\frac{e^{-\varrho t-\sum_{i=1}^2\frac{(x_i-x_{i0}-v_i t)^2}{4D_i t}}}{4\pi t\sqrt{D_1D_2}}\left(1-e^{\frac{(x_{20}-x_{10})(x_1-x_2)}{(D_1+D_2)t}}\right)\,.
\label{eq:c}
\end{equation}
\begin{figure}[h]
\includegraphics[width=5cm,height=5cm]{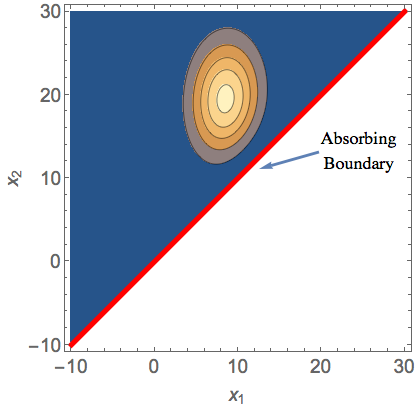}\quad\includegraphics[width=5cm,height=5cm]{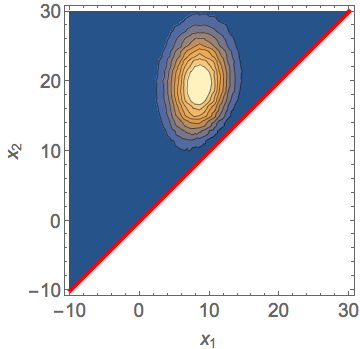}\quad\includegraphics[width=5cm,height=5cm]{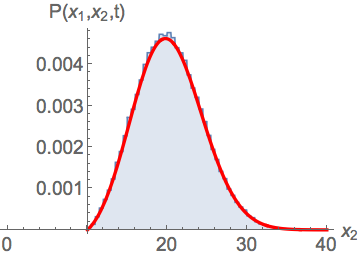}
\caption{ Left: Contour  plot of $P(\bm{x},T)$ for the values   $x_{10}=1$, $x_{20}=2$, $D_1=1$, $D_2=3$, $v_1=2$, $v_2=4$, $T=4$. Middle: Contour plot of the estimate $P(\bm{x},t)$ obtained by Monte Carlo simulations. Right: Comparison of $P(x_1,x_2,T)$ for a fixed value of $x_1=10$ between theory and  Monte Carlo simulations. The plots from the simulations were generated  by using  $10^7$ samples with $\Delta t= 10^{-3}$.}
\label{fig:cp}
\end{figure}
In \fref{fig:cp} we show a density plot of the occupation probability  at time $T=4$ and compared with standard Monte Carlo simulations.\\
From the occupation probability the derivation of the joint first passage probability $\mathcal{F}(x,t)$ at a point $\bm{x}\in\mathcal{H}_x$ at a  time $t$, is proportional to the  probability current crossing the boundary $\mathcal{H}$, that is $\mathcal{F}(x,t)=\phi \left|\left[\left(D_1\partial_{x_1}-D_2\partial_{x_2}\right)P(\bm{x},t)\right]_{x_1=x_2=x}\right|$. In case of having drift and mortality rate,  care  has to be taken as the survival probability may not vanish in the long time limit and a normalisation factor $\phi$ appears. Overall one obtains:
\begin{equation}
\mathcal{F}(x,t)=\phi\frac{|x_{10}-x_{20}|}{4\pi t^2 \sqrt{D_1 D_2}}e^{-\varrho t-\sum_{i=1}^2\frac{ (x-x_{i0}-v_it)^2}{4D_it}}\,.
\label{eq:probc}
\end{equation}
with
\begin{equation}
\phi=e^{\frac{(x_{10}-x_{20})(v_1-v_2)+|x_{10}-x_{20}|\sqrt{(v_1-v_2)^2 +4\varrho( D_1+D_2)}}{2(D_1+D_2)}}\,.
\end{equation}
Notice that the absolute value sign in expression \eref{eq:probc} makes the result valid for $x_{20}>x_{10}$ and $x_{20}<x_{10}$. The expression \eref{eq:probc}  contains all the relevant statistical information of the first passage process. In particular the first passage time probability $\mathcal{F}(t)$ and the eventual hitting probability $\mathcal{E}(x)$ can be obtained by marginalising with respect to $x\in \mathbb{R}$ and to $t\in\mathbb{R}^{+}$, yielding 
\begin{eqnarray}
\mathcal{F}(t)&=&\frac{\delta \eta}{\sqrt{4\pi t^3}}e^{\delta\sqrt{\alpha^2-\beta^2}}e^{-\left(\frac{\alpha^2-\beta^2}{\eta^2}t+\frac{\delta^2\eta^2}{4t}\right)}\, \label{eq:b1},\\
\mathcal{E}(x)&=&\frac{\alpha \delta}{\pi}\frac{ e^{\delta\sqrt{\alpha^2-\beta^2}+\beta\left(x-\mu\right)}}{\sqrt{\delta^2+\left(x-\mu\right)^2}}K_1\left(\alpha\sqrt{\delta^2+\left(x-\mu \right)^2}\right)\,,
\label{eq:b2}
\end{eqnarray}
respectively. Here $K_1(x)$ is the modified Bessel function of the second kind of order $n=1$, while the parameters $(\mu,\delta,\alpha,\beta,\eta)$ are related to the parameters of the problem as follows:
\begin{eqnarray}
\mu&=&\frac{D_2 x_{10}+D_1 x_{20}}{D_1+D_2}\,, \quad\quad \delta=\frac{\sqrt{D_1D_2 }|x_{10}-x_{20}|}{D_1+D_2}\,,\\
\alpha&=&\frac{\sqrt{(D_2 v_1^2+D_1v_2^2+4\varrho D_1D_2)(D_1+D_2)}}{2D_1D_2}\,,\quad\quad \beta=\frac{v_1D_2+v_2D_1}{2D_1 D_2}\,,\\
\eta&=&\frac{D_1+D_2}{D_1D_2}\,.
\end{eqnarray}
The  resulting formula \eref{eq:b2} for the eventual hitting probability is given in terms of the so-called Bessel distribution function, sometimes also known as normal inverse Gaussian distribution function. Using the property that $K_1(x)=\frac{1}{x}+\cdots$ as $x$ tends to $0$, we see that in the limit of zero drift velocity and zero mortality rate we have that
\begin{eqnarray}
\lim_{\bm{v}\,,\varrho\to0} \mathcal{F}(t)&=&\frac{|x_{10}-x_{20}|}{\sqrt{4\pi(D_1+D_2) t^3}}\exp\left(-\frac{(x_{10}-x_{20})^2}{4(D_1+D_2)t}\right)\,,\\
\lim_{\bm{v}\,,\varrho\to0} \mathcal{E}(x)&=&\frac{1}{\pi}\frac{\sqrt{D_1 D_2}(x_{10}-x_{20})}{D_2(x_{10}-x)^2+D_1(x-x_{20})^2}\,.
\end{eqnarray}
In this limit, the expression $\mathcal{E}(x)$ we obtain is a different way of rewriting the result found in\cite{redner2014gradual}.
Notice that formula \eref{eq:b1}, the first passage time probability, is another way of writing the inverse Gaussian with mean $\frac{\delta \eta^2}{2\sqrt{\alpha^2 - \beta^2}}$ and shape parameter $\frac{\delta^2 \eta^2}{2}$.

\section{Statistics properties of the process after $M$ encounters}
\label{sect:ap}
Following \cite{redner2014gradual}  let us suppose that, after the particle is absorbed by the boundary, the process resets to a different initial condition, with different velocities and different diffusion constants, and we repeat this process $n=1,\ldots, M$ times. That is for the $n$-th process for $n=1,\ldots,M$ we have that
\begin{equation}
\mathcal{F}^{(n)}(x,t)=\phi^{(n)}\frac{\left|x^{(n)}_{10}-x^{(n)}_{20}\right|}{4\pi t^2 \sqrt{D^{(n)}_1 D^{(n)}_2}}e^{-\varrho^{(n)}t-\sum_{i=1}^2\frac{\left(x-x^{(n)}_{i0}-v^{(n)}_it\right)^2}{4D^{(n)}_it}}\,.
\label{eq:oe2}
\end{equation}
Then the probability of finding a value of the total time $\tau=\sum_{n=1}^M t_n$ and final point of absorption $\xi=\sum_{n=1}^M x_n$ after $M$ encounters is given by the convolution of \eref{eq:oe2} over the whole process, that is
\begin{equation}
\hspace{-2cm}\mathcal{F}_{M}(\xi,\tau)=\int_0^\infty\int _{-\infty}^{\infty} \left[\prod_{n=1}^M dx_n dt_n\mathcal{F}^{(n)}_{\mathcal{H}}(x_n,t_n)\right]\delta\left(\tau-\sum_{n=1}^M t_n\right)\delta\left(\xi-\sum_{n=1}^M x_n\right)\,.
\end{equation}
In the Laplace-Fourier space the convolution becomes a function product of the Laplace-Fourier transform of the joint probability density function \eref{eq:oe2}:
\begin{equation}
\widehat{\mathcal{F}}_{M}(k,s)=\int_0^\infty d\tau e^{-s\tau} \int_{-\infty}^\infty d \xi e^{ik\xi}\mathcal{F}_{M}(\xi,\tau)=\prod_{n=1}^M \widehat{\mathcal{F}}^{(n)}(k,s)\,,
\end{equation}
with $\widehat{\mathcal{F}}^{(n)}_{\mathcal{H}}(k,s)$ being the Fourier-Laplace transform of \eref{eq:oe2} with $(\mu^{(n)},\delta^{(n)},\alpha^{(n)},\beta^{(n)})$ parameters and given by
\begin{equation}
\widehat{\mathcal{F}}^{(n)}_{\mathcal{H}}(k,s)=\exp\left[i k \mu+\delta\sqrt{\alpha^2-\beta^2}-\delta\sqrt{\alpha^2-(\beta+ik)^2+\eta^2 s}\right]\,.
\label{eq:a}
\end{equation}
For a general sequence of parameters $\{(\mu^{(n)},\delta^{(n)},\alpha^{(n)},\beta^{(n)})\}_{n=1}^M$ the aggregate eventual hitting probability will usually tend to a Gaussian distribution function. Here we look at  particular sequences for which there exists an stable distribution. Looking at \eref{eq:a} for $s=0$, we notice that the Bessel distribution function only becomes stable under convolution when the parameters $\alpha$ and $\beta$ are kept constant (see for instance \cite{Anna}), and thus, the  eventual hitting probability after $M$ encounters will also be a Bessel distribution.\\
With this in mind, we study a resetting process that obeys the following two constraints
\begin{eqnarray}
\hspace{-1cm}4\alpha^2&=&\left( \frac{(v^{(n)}_1)^2}{D^{(n)}_1}+\frac{(v_2^{(n)})^2}{D^{(n)}_2}+4\varrho^{(n)}\right)\left(\frac{1}{D^{(n)}_1}+\frac{1}{D^{(n)}_2}\right)\,,\quad n=1,\ldots,M\,,\label{eq:c1}\\
\hspace{-1cm}2\beta&=&\frac{v^{(n)}_1}{D^{(n)}_1}+\frac{v^{(n)}_2}{D^{(n)}_2}\,,\quad n=1,\ldots,M\,.\label{eq:c2}
\end{eqnarray}
\begin{figure}
\centering
\includegraphics[width=7cm, height=6cm]{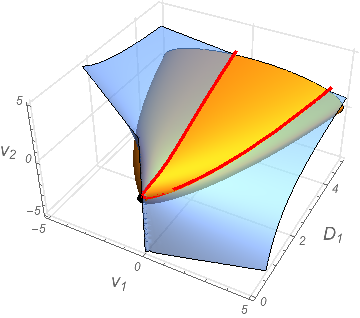}\quad\quad\includegraphics[width=7cm, height=6cm]{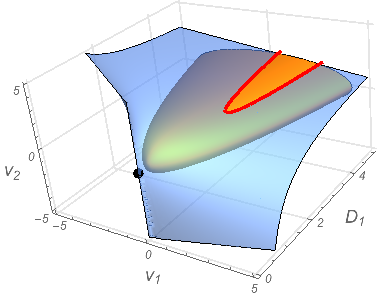}
\caption{Plot of the set of constraints \eref{eq:c1} and \eref{eq:c2} for $\alpha=1.2$, $\beta=1$ and $D_2=1$ and $\varrho=0$ (left plot)  or $\varrho=0.3$ (right plot). The intersection of the two constraints is shown with a red thick line. Any process that follows these lines has a stable  Bessel distribution. In the context of the man-mosquitoes problem, the black point  at coordinates $(v_1,D_1,v_2)=(0,0,0)$ corresponds to the case of the man captured by a satiated mosquito.}
\label{fig:param3D}
\end{figure}
In figure \ref{fig:param3D} we show a plot of the set of constraints \eref{eq:c1} and \eref{eq:c2}. Their intersection corresponds to the parameters-line along which the convolution of the Bessel distribution function describing the eventual hitting probability is stable.\\
In the context of the man-mosquito problem, we wonder whether there exists a subsequence of encounters in the space of parameters for which the Bessel distribution is stable and the mosquito achieves to slow down the man, perhaps to his death. As we can see from the set of equations \eref{eq:c1} and \eref{eq:c2} this is certainly possible. Indeed, let us recall that in the context of the man-mosquitoes problem, the first particle represents the man  and the second particle represents the mosquito.  Man's death after $M$ encounters corresponds to having $v_1^{(n)}=D_1^{(n)}=0$ for $n=M$. In order to achieve this and at the same time satisfy the stability constraints \eref{eq:c1} and \eref{eq:c2}, the mortality rate must either be zero or evolve to zero, while the drift of the mosquito must come to a halt  with its diffusion constant being different from zero, that is $v_{2}^{(n)}=0$ and $D_2^{(n)}\neq 0$ for $n=M$. Here one can say that once the mosquito captures the man, the mosquito is satiated. Alternatively, if the mortality rate is different from zero, the mosquito may manage to gradually slow down the man after $M$ encounters, that is $v_1^{(n)}=0$ ,$D_1^{(n)}\neq 0$ for $n=M$, but is unable to capture him.
\begin{figure}
\begin{center}
\includegraphics[width=7cm,height=5cm]{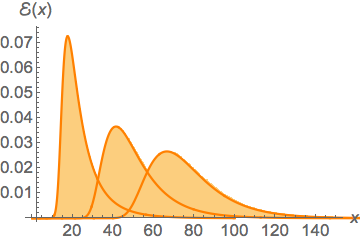}\quad\quad\includegraphics[width=7cm,height=5cm]{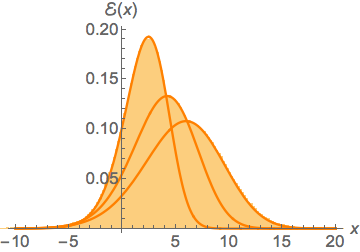}
\end{center}
\caption{Left plot: eventual hitting probability $\mathcal{E}_{M}(x)$ for $M=3$ (leftmost), $6$ and $9$ (rightmost) encounters. Here we compare the analytical results (thick orange lines) with those obtained from Monte Carlo Simulations using $10^6$ samples (orange filling). Right plot:  eventual hitting probability with mortality rate for $M=3$ (leftmost), $6$ and $9$ (rightmost) encounters. The evolution line of the parameters obeys the constraints \eref{eq:c1} and \eref{eq:c2} with $\alpha=1.0606$, $\beta=1$, and $D_2=1$, and $\varrho=0$ (left plot) and   $\alpha=2.71109$, $\beta=-3/2$, and $D_2=1$, and $\varrho=1/10$ (right plot).}
\label{fig:af}
\end{figure}

\section{Monte Carlo Simulations}
\label{sect:mcs}
To check the validity of our analytical we have performed  thorough Monte Carlo simulations. In the standard approach, let us consider a $2$-dimensional BM $\bm{x}(t)=(x_1(t), x_2(t))$. with drift vector $\bm{v}=(v_1, v_2)$. We have used  a simple Euler discretisation to write the  Langevin equation as follows.
\begin{equation}
x_i(t+\Delta t)=x_i(t)+v_i\Delta t+\sqrt{2D_i \Delta t}\mathcal{N}_{i}(0,1)\,,\quad\quad i=1,2
\label{eq:ed}
\end{equation}
with $\{\mathcal{N}_{i}(0,1)\}_{i=1,2}$ are two independent Gaussian variables with zero mean and unity standard deviation. In case of having a nonzero  mortality rate $\varrho$, one simply chooses to remove one particle from the ensemble with probability $\varrho\Delta t$.  From here we can get estimates for the occupation probability $P(\bm{x},t)$   and the joint first passage probability $\mathcal{F}(x,t)$ given by equations \eref{eq:c} and \eref{eq:probc}, respectively.\\
To obtain an estimate of the occupation probability we prepare an ensemble of $\mathcal{N}_{{\rm ens}}$ particles with a deterministic initial condition $x_{1,a}(t_0)=x_{10}$ and $x_{2,a}(t_0)=x_{20}$ for $a=1,\ldots,\mathcal{N}_{{\rm ens}}$ and with $x_{20}>x_{10}$. The particles of the ensemble are iterated according to eqs. \eref{eq:ed} and if $x_2(t)>x_1(t)$. If a given particle crosses the boundary $x_1=x_2$ then it becomes inactive/absorbed and it is no longer iterated. This process is continued up to a time $T$ at which we want to get an histogram  $P(\bm{x},T)$. If we denote as $\{\bm{x}_{a}\}_{a=1}^{\mathcal{N}_{{\rm act}}}$ the set of coordinates of those particles that remained active up to time $T$ then $P(\bm{x},T)\sim (\mathcal{N}_{{\rm act}}/\mathcal{N}_{{\rm ens}})\sum_{a=1}^{\mathcal{N}_{{\rm act}}}\delta(\bm{x}-\bm{x}_a)$. This is the method used to generate the estimates that appeared in \fref{fig:cp}.\\
Similarly, to obtain an estimate of $\mathcal{F}(t,x)$, we start again with an initial condition $x_{20}>x_{10}$ and we simply iterate the  equations \eref{eq:ed} until  the particle crosses the boundary $x_1=x_2$. The pair  $(t,x)$ with $x=x_1=x_2$ is recorded, corresponding to the time $t$ of absorption and the position $x$ along the boundary at which the particle is absorbed. To obtain a histogram of $\mathcal{F}(t,x)$ we run this process over an ensemble of $\mathcal{N}_{{\rm ens}}$ particles, so that we have a collection of pairs $\{(t_{a},x_a)\}_{a=1}^{\mathcal{N}_{{\rm ens}}}$ from which we can estimate $\mathcal{F}(t,x)\sim (1/\mathcal{N}_{{\rm ens}})\sum_{a=1}^{\mathcal{N}_{{\rm ens}}}\delta(t-t_a)\delta(x-x_a)$. For the case of $M$ encounters we simply run this approach separately for each encounter with its corresponding parameters and then construct the histogram of $\mathcal{F}_M(t,x)$ of the aggregate process\footnote{Depending on the value of the parameters, it will happen somewhat frequently that the particle drifts  in the opposite direction to the absorbing boundary. To avoid  the  Monte Carlo simulation to slow down in these cases, we just include a time cutoff for the particle to be absorbed by the boundary.}.\\
In figure \ref{fig:af} we show a comparison between the theory of Monte Carlo simulations of the eventual hitting probability $\mathcal{E}_M(x)$ after $M$ encounters, for a process that obeys the set of constraints \eref{eq:c1} and \eref{eq:c2} and for which the man is captured by a satiated mosquito. This process was   performed by using the values $\alpha=1.0606$, $\beta=1$ and $D_2=1$.

\section{Summary and outlooks}
\label{sect:c}

In this work we have studied the first passage time properties of a  two-dimensional  particle with anisotropic diffusion, mortality rate and a drift velocity, in the presence of an absorbing boundary. After the particle is absorbed at the boundary, the process is restarted with updated values of its diffusion constants and drift velocity. The resulting joint probability that, after $M$ encounters, the particle is absorbed at a point  of the boundary  at a given time is then derived and we have shown that under certain conditions the  eventual hitting probability follows a Bessel distribution.\\
Thus the present work generalises the literature concerning prey-predator one-dimensional problems in non-biased scenarios by adding two basic ingredients: bias  and a mortality rate. More specifically in the context of the man-mosquitoes problem, the mosquito is able to gradually capture the man, after which the mosquito follows a diffusion process with no drift. Our results are compared thoroughly with Monte Carlo simulations showing excellent agreement.

\ack
The authors  thank D. Boyer, I. Pineda, F. Font-Clos and S. Redner for discussions and critical reading of the manuscript.

\section*{References}

\bibliographystyle{iopart-num}

\bibliography{mybib}

\end{document}